\documentclass[12pt]{article}
\usepackage{amssymb}
\usepackage{setspace}

\evensidemargin \oddsidemargin
\def\b#1{{\mathbb #1}}

\author{Gaetano Fiore\\
Dip. di Matematica e Applicazioni,\\
Universit\`a Federico II, v. Claudio 21, 80125 Napoli,\\
and I.N.F.N., Sezione di Napoli, Italy
\\ \\
Laure Gouba \\
The Abdus Salam International Centre for
Theoretical Physics (ICTP),\\
 Strada Costiera 11,
I-34151 Trieste Italy \\
Email: lgouba@ictp.it}
\begin{title}
{ Class of invariants for the 2D time-dependent
Landau problem and harmonic oscillator
in a magnetic field}
\end{title}
\begin{document}
\maketitle
\abstract{We consider an isotropic two dimensional harmonic oscillator with
arbitrarily time-dependent mass $M(t)$ and frequency $\Omega(t)$ in
an arbitrarily time-dependent magnetic field $B(t)$. We determine 
two commuting invariant
observables (in the sense of Lewis and Riesenfeld) $L,I$
in terms of some solution of an auxiliary ordinary
differential equation and an orthonormal basis of the Hilbert space
consisting of joint eigenvectors $\varphi_\lambda$ of  $L,I$. We
then determine time-dependent phases $\alpha_\lambda(t)$ such that
the $\psi_\lambda(t)=e^{i\alpha_\lambda}\varphi_\lambda$ are
solutions of the time-dependent Schr\"odinger equation and make up
an orthonormal basis of the Hilbert space. These results apply, 
in particular to a two dimensional Landau problem with
time-dependent $M,B$, which is obtained from the above just by setting
$\Omega(t)\!\equiv\! 0$. By a mere redefinition of
the parameters, these results can be applied also to the analogous
models on the canonical non-commutative plane.}

\section{Introduction}

The study of time-dependent quantum problems
has been of great interest in the literature since the work of  Lewis
and Riesenfeld \cite{lewis,LewRie69},  in the late $1960$s.
They developed a completely general  approach to the evolution
generated by a time-dependent Hamiltonian $H(t)$;
one re-obtains the results of the adiabatic
approximation in the case of slowly varying $H(t)$, and the results
of the ``sudden'' approximation in the case of  fast or even
discontinuously varying $H(t)$. In  Refs.\cite{lewis} and \cite{LewRie69},
the method was applied to a one-dimensional harmonic oscillator
with a $t$-dependent frequency and to a charged particle
in a $t$-dependent, axially symmetric electromagnetic field in
three dimensions. For zero electric field, the latter reduces
to a Landau problem with time-dependent magnetic field.
Several other applications of the method
have been worked out later; we just mention the application in 
 \cite{GueLea77}
to the anisotropic three-dimensional harmonic oscillator, the ones
\cite{Lea78} to the damped harmonic oscillator and to a
\cite{LuaTan05} linear (in space) potential in one dimension.
In the present work, we apply the approach of Lewis and
Riesenfeld to a two-dimensional isotropic harmonic oscillator
with time-dependent mass $M$ and frequency $\Omega$
in a time-dependent (but space-independent
magnetic field) $B$ [formula (\ref{Mt})] [the term
$\Omega^2 x\!\cdot\! x$ in the Hamiltonian may contain,
in particular, an electric potential of the form
$\varphi(x)=\frac{e\eta(t)}{2M(t)c^2}x\!\cdot\! x$
\cite{Chee09}, \cite{footnote}]
for $\Omega\equiv 0$, 
this reduces to a Landau problem with time-dependent mass $M,B$.
In the Coulomb gauge, the Hamiltonian can be rewritten as
the sum of the Hamiltonian of a harmonic oscillator
with suitable time-dependent mass $\mu$ and frequency $\omega$
and of the angular momentum $L$ multiplied by a suitable
time-dependent  coefficient $\nu$ [formula (\ref{hmt})].
In either case the model can be immediately embedded in a
three-dimensional one where $B$ is directed along the third direction, and
the motion in the direction of the latter is free; the latter
would describe, in particular, a charged particle in a suitable $t$-dependent
axially symmetric electromagnetic field (as in \cite{LewRie69}, but with a possibly
$t$-dependent mass).

The results can be applied also to a two dimensional
isotropic harmonic oscillator with time-dependent $M,\Omega,B$ on the
(canonical) {\it non-commutative} plane [formula (\ref{hMt})]. In fact,
the Hamiltonian can be again rewritten in the same form (\ref{hmt})
with new parameters $\mu,\omega,\nu$ involving also the deformation 
parameter $\theta$ of the non-commutative plane in their definition.
Non-commutative space-time structure is an old subject dating back
over fifty years to Snyder \cite{sny}. The analysis of field theories
on non-commutative space-times has become a promising area
in theoretical physics \cite{doug}, and also  non-commutative quantum mechanics
has become subject for intense investigations. In the recent works \cite{laure}
the formulation, interpretation and applications of non-commutative quantum
mechanics have been investigated in configuration space, with the explicit
example of the (time-independent) two-dimensional harmonic oscillator
in \cite{laure}, or general Hamiltonians (including
the one of the Landau problem) in \cite{Fio10}, where an approach
to second quantization is also proposed.

In section $2$, we define the model and use the socalled Bopp-shift
to connect the noncommutative variables to the commutative ones.
In section $3$, we determine two commuting invariant observables $L,I$
($L$ is the angular momentum) in terms of
some solution of an auxiliary nonlinear ordinary differential
equation. In section $4$, we determine
an orthonormal basis of the Hilbert space consisting of joint eigenvectors of  $L,I$;
the presence of the parameter of noncommutativity $\theta$
does not modify much the results.
Then, we determine suitable time-dependent phase factors such that
their products with these joint eigenvectors make up
an orthonormal basis of the Hilbert space consisting of
solutions of the Schr\"odinger equation. Each phase factor
can be split as the product  of a  dynamical one and
geometric one (including the Berry phase) \cite{Berry84 - AhaAna87}.
The latter includes, in general, the effects of the socalled adiabatic
operator; the latter multiplied by the socalled dynamical operator
and the non-adiabatic one gives a decomposition of the evolution operator,
as explained e.g., in the introduction of \cite{Chee10}
(see also \cite{Messiah}).

\medskip
In the rest of this introduction,
we briefly recall the general Lewis-Riesenfeld approach to time-dependent
quantum mechanics \cite{LewRie69}. An operator
$I(t)$ is said to be an invariant if
\begin{eqnarray}\label{eq}
 \dot I(t) \equiv \frac{\partial}{\partial t} I(t) +
\frac{1}{i\hbar}\left[I(t), H(t)\right] = 0
\end{eqnarray}
(note that $H$ itself is not an invariant unless it is time-independent).
Here and in the sequel we use the Schr\"odinger ({\it not} the Heisenberg)
picture.  If $I(t)$  is  an invariant and
$\psi(t)$ solves the Schr\"odinger equation
\begin{equation}\label{Schr}
i\hbar\frac{\partial\psi}{\partial t} (t)=H(t)\psi(t),
\end{equation}
then also $I(t)\psi(t)$ does. If in addition $I(t)$ is hermitian,
then there exists a (in general time-dependent) orthonormal basis
$\{\varphi_{\lambda,\kappa}\}$ of
eigenvectors of $I$ with real eigenvalues $\lambda\in\Lambda$;
these eigenvectors are parameterized by  $\lambda\in\Lambda$
and if necessary  by some additional label $\kappa\in K$.
Deriving the relation $I\varphi_{\lambda,\kappa}=\lambda\varphi_{\lambda,\kappa}$
with respect to $t$ and using (\ref{eq}),
(\ref{Schr}) one finds that all eigenvalues $\lambda$ are
time-independent. Moreover, adjusting the phase, the $\varphi_{\lambda,\kappa}$
can be chosen so as to be a complete set of solutions of (\ref{Schr}),
provided $I$ does not contain the operator $\partial_t$
(for instance, operators of the form $I=\alpha\partial_t$ are not allowed).
If $H$ is time-independent and we choose
a time-independent $I$, then eq. (\ref{eq})
amounts to $[I,H]=0$, and
the above scheme amounts to finding a complete set of
solutions of  (\ref{Schr}) in the form of eigenvectors
of both $I$ and $H$, as usual.

\section{The model}

The Hamiltonian for the two dimensional isotropic harmonic oscillator
in a  magnetic field reads
\begin{eqnarray}\label{Mt}
 H_{hm}(t) &=& \frac{1}{2M(t)}\,[p\!-\!eA(t)]\cdot[p\!-\!eA(t)]
+\frac{M(t)\Omega^2(t)}{2}\, x\cdot x\\
 &=&\frac{1}{2M(t)}\left[\left(p_1\!+\!\frac e2
B(t)x_2\right)\!^2 + \left(p_2\!-\!\frac e2 B(t)x_1\right)\!^2\right]
+\frac{M(t)\Omega^2(t)}2\left( x^2_1 + x^2_2\right);\nonumber
\end{eqnarray}
we assume piecewise
continuously time-dependent mass $M$, frequency $\Omega$
and but magnetic field $B$ (assumed space-independent).
The second line holds after expressing the vector potential in
Coulomb gauge. As said, for $\Omega\equiv 0$
this reduces to a Landau problem with time-dependent mass $M,B$,
\begin{equation}
 H_L(t) = \frac{[p\!-\!eA(t)]\cdot[p\!-\!eA(t)]}{2M(t)}
=\frac{1}{2M(t)}\left[\left(p_1\!+\!\frac e2
B(t)x_2\right)^2 + \left(p_2\!-\!\frac e2 B(t)x_1\right)^2\right].
\end{equation}
We use units such that $\hbar = 1$ and $x_a,p_a,\omega$ are
dimensionless
(this can be achieved renormalizing as usual the latter variables
by dimensionful factors built out of
some constant mass $m_0$ and frequency $\omega_0$, and $\hbar$).
Using the canonical commutation relations
$[x_a,x_b]=0=[p_a,p_b]$, $[x_a,p_b]=i\delta_{ab}$
it is straightforwad to check that $H_{hm}(t)$ reduces to
\begin{eqnarray}\nonumber
 H(t) &=& \frac{1}{2\mu(t)}\,p\cdot p
+\frac{\omega^2(t)}{2}\, x\cdot x- \frac{\nu(t)}{2}L \\
&=&\frac{1}{2\mu(t)}\left(p_1^2 + p_2^2\right)
+\frac{\omega^2(t)}2\left( x^2_1 + x^2_2\right)- \nu(t)L,\label{hmt}
\end{eqnarray}
where $L = x_1p_2 -x_2p_1$ is the angular momentum, and
\begin{equation}
\mu(t):=M(t), \qquad \omega^2(t):=M(t)\Omega^2(t)\!+\!
\frac {[eB(t)]^2}{4M(t)}, \qquad \nu(t):=\frac {eB(t)}{2M(t)}.
\label{parameters1}
\end{equation}

Our two-dimensional non-commutative space is defined by
noncommuting coordinates   $\hat X_1, \hat X_2$
and associated momentum components $\hat P_1, \hat P_2$ satisfying
the following commutation relations:
\begin{eqnarray}
 \left[\hat X_1,\hat X_2 \right] = i\theta,\quad
\left[\hat X_1,\hat P_1 \right]  = i,\quad
\left[\hat X_2,\hat P_2 \right]  = i,\quad
\left[\hat P_1,\hat P_2 \right]  = 0.
\end{eqnarray}
All these operators are hermitian. On such a non-commutative space
we consider as a Hamiltonian that of a two dimensional  isotropic harmonic oscillator
in a  magnetic field with $t$-dependent $M,\Omega,B$,
\begin{equation}\label{hMt}
\hat H_{hm}(t) =\frac{1}{2M(t)}\left[\left(\hat P_1\!+\!\frac e2
B(t)\hat X_2\right)\!^2 + \left(\hat P_2\!-\!
\frac e2 B(t)\hat X_1\right)\!^2\right]
+\frac{M(t)\Omega^2(t)}2\left( \hat X^2_1 + \hat X^2_2\right).
\end{equation}
The aim is to find two independent invariants and then solve
the Schr\"odinger equation within each eigenspace of the latter.
We redefine the operators as follows
\begin{eqnarray}
x_1 = \hat X_1 +\frac{\theta}{2}\hat P_2,\qquad
x_2 = \hat X_2 -\frac{\theta}{2}\hat P_1,\qquad
p_1 = \hat P_1,\qquad
p_2 = \hat P_2;
\end{eqnarray}
the associated commutations relations are
\begin{eqnarray}
\left[x_1, x_2\right] = 0,\qquad
\left[x_1, p_1\right] = i,\qquad
\left[x_2, p_2\right] = i,\qquad
\left[p_1, p_2\right] = 0.
\end{eqnarray}
Such a redefinition is sometimes also called the Bopp shift. In
terms of these new operators $\hat H_{hm}(t) $
reduces again to (\ref{hmt}), where now
\begin{equation}
\mu^{-1}:=\frac {[4\!+\!eB\theta]^2}{16M}\!+\!M\Omega^2\frac {\theta^2}{4},
\qquad \omega^2:=M\Omega^2\!+\!
\frac {[eB]^2}{4M}, \qquad -\nu:=\frac {[4\!+\!eB\theta]eB}{8M}
\!+\!\frac {M\Omega^2\theta}2.
\end{equation}
(we have omitted the time dependence for brevity).
Note that in general $\mu,\omega$ are $t$-independent only if all of
$M,B,\Omega$ are.

Therefore for $B(t)\equiv 0$, we obtain the harmonic oscillator
with time-dependent frequency and mass, while
for $\Omega(t)\equiv 0$ we obtain the Landau problem with time-dependent 
frequency and mass, both on the commutative and noncommutative plane.

An operator invariant of the model (\ref{hmt}) is an operator
that satisfies eq. (\ref{eq}).

\section{Derivation of the operator invariants}

As $x\!\cdot\! x,\, p\!\cdot\! p,\,x\!\cdot\! p,\,p\!\cdot\! x$,
and therefore $H$, are explicitly symmetric under rotations,
they commute  with the generator of rotations $L$,
\begin{equation}
[L,x\!\cdot\! x]=0,\quad [L,p\!\cdot\! p]=0,\quad
[L,x\!\cdot\! p]=0,\quad [L,p\!\cdot\! x]=0,\quad
[L,H]=0.                         \label{rotsym}
\end{equation}
Since $L$  is also time-independent, it fulfills
(\ref{eq}) and is a first operator
invariant. In order to find another one $I(t)$
commuting with $L$, we make the Ansatz
\begin{eqnarray}
\nonumber\label{eq11}
I(t) &=& \alpha(t)\,x\cdot x
+\beta(t)\,p\cdot p
+\gamma(t)\left(x\cdot p+p\cdot x \right)+\delta(t)L\\[8pt]
&=&\alpha(t)\left( x_1^2 +  x_2^2\right)
+\beta(t)\left( p_1^2 + p_2^2 \right)
+\gamma(t)\left(\{x_1,p_1\}
+ \{x_2,p_2\}\right)+\delta(t)L,
\end{eqnarray}
where $\{x_1,p_1\} = x_1 p_1 + p_1 x_1$, $\{x_2,p_2\} =
x_2 p_2 + p_2 x_2$.
As we need $I(t)$ to be hermitean, $I^\dagger = I$, the
coefficients $\alpha,\:\beta,\gamma,\delta$
 should be real. As said,
\begin{equation} \label{ILcom}
 [I, L] = 0,
\end{equation}
because $I$ is manifestly symmetric under
rotations. Using (\ref{rotsym}) and
$$
\begin{array}{llll}
\left[x_1^2, p_1^2 \right] = 2i\{ x_1, p_1 \},\qquad &
\left[x_1^2, \{ x_1, p_1 \} \right] = 4i x_1^2,\qquad &
\left[p_1^2, \{ x_1, p_1 \} \right] = -4i p_1^2, \\[8pt]
\left[x_2^2, p_2^2 \right] = 2i\{ x_2, p_2 \},\qquad  &
\left[x_2^2, \{ x_2, p_2 \} \right] = 4i x_2^2,\qquad &
\left[p_2^2, \{ x_2, p_2 \} \right] = -4i p_2^2,
\end{array}
$$
Eq. (\ref{eq}) reduces to
\begin{eqnarray}\label{eq1}
\dot \alpha - 2\omega^2\gamma  &=& 0,\\\label{eq2} \dot \beta +
\frac 2\mu \gamma &=& 0,\\\label{eq3} \dot \gamma + \frac 1\mu \alpha - \omega^2
\beta  &=& 0,\\\label{delta} \dot\delta &=& 0.
\end{eqnarray}
Eq. (\ref{delta})
implies that $\delta$ is a constant.
The solution of the system (\ref{eq1})-(\ref{delta})
with $\delta=1$ and all the other coefficients vanishing
is the already determined invariant $L$.
In the rest of the text, we choose $\delta = 0$ for the other invariant.
From (\ref{eq1})-(\ref{eq3}) it follows
\begin{equation}
 \frac{d}{dt}\left(\gamma^2 -\alpha\beta \right) = 0,
\end{equation}
whence
\begin{equation}\label{eq13}
 \gamma^2 - \alpha\beta = -\kappa^2,
\end{equation}
where $\kappa$ is either a real or an imaginary constant
(so that $\kappa^2$ is real).

\medskip
Since $I(t)$ and $L$ commute, they have joint eigenvectors.
One can find all of the latter using the same
operator technique as in Ref. \cite{lewis}, that is the Dirac method of
diagonalizing the Hamiltonian of a constant-frequency harmonic oscillator.
To this end, we choose a positive $\kappa$,
stick to positive-definite solutions $\beta(t)$ and
introduce an auxiliary real-valued function $\sigma(t)$ such that
\begin{equation}\label{eq5}
\beta (t) = \sigma^2(t);
\end{equation}
then by Eqs. (\ref{eq2}) and (\ref{eq13})
\begin{equation}\label{eq6}
 \gamma = -\mu\sigma\dot\sigma,
\qquad  \qquad
 \alpha = \dot\sigma^2 \mu^2 +\frac{\kappa^2}{\sigma^2}.
\end{equation}
Replacing (\ref{eq5}), (\ref{eq6}) in
(\ref{eq3}), we find that the system (\ref{eq1}-\ref{eq3})
is equivalent to the auxiliary equation
\begin{eqnarray}\label{neq1}
 \ddot{\sigma} +\frac{\dot \mu}{\mu}\dot\sigma
+ \frac{\omega^2}{\mu}\sigma =  \frac{\kappa^2}{\mu^2\sigma^3}.
\end{eqnarray}
In terms of a real solution $\sigma(t)$ of (\ref{neq1}) the  operator
invariant (\ref{eq11}) takes the form
\begin{eqnarray}\label{eq14}
I (t) = \left( \dot\sigma\mu x_1 -\sigma p_1\right)^2
+\frac{\kappa^2}{\sigma^2} x_1^2 + \left( \dot\sigma\mu
x_2 -\sigma p_2\right)^2 +\frac{\kappa^2}{\sigma^2} x_2^2.
\end{eqnarray}

\section{Eigenstates of $I(t)$ and the phases}

We can introduce ``raising'' and ``lowering '' operators
$a_1,\: a_2,\: a_1^\dagger,\: a_2^\dagger$ by
\begin{eqnarray}
 a_a(t) = \frac{1}{\sqrt{2\kappa}}
\left(\dot\sigma\mu x_a -\sigma p_a +
i\frac{\kappa}{\sigma} x_a\right), \qquad \quad a_a^\dagger(t) =
\frac{1}{\sqrt{2\kappa}}  \left(\dot\sigma\mu x_a -
\sigma p_a - i\frac{\kappa}{\sigma} x_a\right),\label{g1}
\end{eqnarray}
with $a=1,2$.
They fulfill the commutation relations
\begin{eqnarray}
 [a_1,\:a_1^\dagger] = 1,\quad [a_2,\:a_2^\dagger] = 1,\quad
[a_1,\: a_2] = 0,\quad [a_1,\:a_2^\dagger] = 0.
\end{eqnarray}
Then the operator invariants $I,L$ can be written as
\begin{equation}
 I_{-}(t) = 2\kappa\left(a_1^\dagger a_1  +
 a_2^\dagger a_2 + 1\right),
\qquad \qquad L = i(a_2^\dagger a_1 - a_1^\dagger a_2).
\end{equation}
Instead of working with the $a_a,a_a^\dagger$ it is convenient
to work with  the left and right circular annihilation operators
$A_{-}, A_{+}$, that are
respectively defined by
\begin{equation}
 A_{-}(t) = \frac{1}{\sqrt{2}}\left(a_1 -i a_2 \right),\qquad\qquad
A_{+}(t) = \frac{1}{\sqrt{2}}\left(a_1 + i a_2 \right),
\end{equation}
and the left and right creation operators, that are the
hermitean conjugates of the above:
\begin{equation}\label{fin0}
 A_-^\dagger(t) = \frac{1}{\sqrt{2}} \left(a_1^\dagger +i a_2^\dagger \right),\qquad
\qquad A_+^\dagger(t) = \frac{1}{\sqrt{2}} \left(a_1^\dagger -i
a_2^\dagger \right).
\end{equation}
The only non-zero commutators among $A_{-},\, A_{+},\;
A_+^\dagger,\; A_-^\dagger$
 are then given by
\begin{equation}
 [A_{-}, A_-^\dagger] = 1,\qquad \qquad [A_{+}, A_+^\dagger] = 1.
\end{equation}
Using the inverse relations
\begin{eqnarray}\nonumber
 a_1 &=& \frac{\sqrt{2}}{2}(A_{-} + A_{+}),\qquad\qquad
a_1^\dagger = \frac{\sqrt{2}}{2}(A_+^\dagger + A_-^\dagger),\\
\nonumber a_2 &=& \frac{i\sqrt{2}}{2}(A_{-} - A_{+}),\qquad\qquad a_2^\dagger
= \frac{i\sqrt{2}}{2}(A_+^\dagger - A_-^\dagger),
\end{eqnarray}
we find
\begin{equation}
\begin{array}{ll}
 [I_{-},A_{\pm}] = -2\kappa,\qquad \qquad &[I_{-},A_{\pm}^\dagger] = 2\kappa,\\[8pt]
[L,A_{\pm}] = \pm A_{\pm},\qquad \qquad &[L,A_{\pm}^\dagger] = \mp A_{\pm}^\dagger,
\end{array}              \label{bla}
\end{equation}
and
\begin{equation}
 I_{-}(t) = 2\kappa (A_+^\dagger A_{+} + A_-^\dagger A_{-} + 1),
\qquad \qquad L = (A_-^\dagger A_{-} - A_+^\dagger A_{+}).
\end{equation}
Let us assume $\Vert 0,0 \rangle$  is a normalized state
annihilated by $a_1, a_2$, or equivalently by $A_\pm$:
\begin{equation}
 A_{-}\Vert 0,0 \rangle = 0, \qquad \qquad A_{+}\Vert 0,0 \rangle = 0.
\end{equation}
For any  $r_{+},\; r_{-}\in\b{N}_0$ ($\b{N}_0$ stands for
the set of nonnegative integers) we find
\begin{equation}
\begin{array}{l}
 I_{-}(t)(A_-^\dagger)^{r_{+}}(A_+^\dagger)^{r_{-}}\Vert 0,0 \rangle =
2\kappa (r_{+} + r_{-} + 1) (A_-^\dagger)^{r_{+}}(A_+^\dagger)^{r_{-}}\Vert 0,0 \rangle ,\\[6pt]
L(t)(A_-^\dagger)^{r_{+}}(A_+^\dagger)^{r_{-}}\Vert 0,0 \rangle =
(r_{+} -r_{-}) (A_-^\dagger)^{r_{+}}(A_+^\dagger)^{r_{-}}\Vert 0,0 \rangle.
\end{array}              \label{eigen}
\end{equation}
The pairs $(r_{+}, r_{-})$ are in one-to-one correspondence with the
pairs $(n,m)$, with $n,m$ defined by
\begin{equation}
n = r_{+} + r_{-}\in\b{N}_0,\qquad\qquad
m = r_{+} -r_{-}\in \{-n,2\!-\!n,...,n\!-\!2,n\}.   \label{defn}
\end{equation}
The inverse of (\ref{defn}) are
\begin{equation}
r_{+} = \frac{n+m}2 \in\b{N}_0,\qquad\qquad r_{-} =
\frac{n-m}2 \in\b{N}_0.             \label{invdefn}
\end{equation}
An orthonormal basis of the Hilbert space of states
of the system consists of
\begin{equation}
 \Vert  n,m \rangle = \mathcal{N}
(A_-^\dagger)^{\frac{1}{2}(n+m) }
(A_+^\dagger)^{\frac{1}{2}(n-m)}\Vert 0,0 \rangle ,
\qquad\quad
\mathcal{N}= \frac{1}{\sqrt{ \frac{n\!+\!m}2!}}\frac{1}{\sqrt{ \frac{n\!-\!m}2!}}
\end{equation}
which by  (\ref{eigen}) are eigenvectors of $I_{-}(t)$  and  $L$:
\begin{equation}
I \Vert  n,m \rangle =2\kappa (n+1) \Vert  n,m \rangle,\qquad\qquad
L\Vert  n,m \rangle =m \Vert  n,m \rangle. \label{eigenvalues}
\end{equation}
The action of $A_\pm,A_\pm^\dagger$ on the basis reads:
\begin{eqnarray}
&& A_+\Vert  n,m \rangle = \sqrt{\frac{n\!+\!m}2}\:
\Vert  n\!-\!1,m\!-\!1 \rangle,\qquad\quad
A_+^\dagger\Vert  n,m \rangle=\sqrt{\frac{n\!+\!m}2\!+\!1}\:
\Vert  n\!+\!1,m\!+\!1 \rangle,\\
&& A_-\Vert  n,m \rangle = \sqrt{\frac{n\!-\!m}2}\:
\Vert  n\!-\!1,m\!+\!1 \rangle,\qquad\quad
A_-^\dagger\Vert  n,m \rangle=
\sqrt{\frac{n\!-\!m}2\!+\!1}\: \Vert  n\!+\!1,m\!-\!1 \rangle.
\end{eqnarray}

By the general theory \cite{LewRie69}, one can transform the basis $\{\Vert  n,m \rangle\}$
into an orthonormal basis $\{ \psi_{n,m}\}$ consisting of solutions of the
equation  (\ref{Schr}) applying  suitable phase transformations
\begin{equation}
 \psi_{n,m} =  e^{i\alpha_{m,n}(t)}\Vert  n,m \rangle
\end{equation}
By the orthonormality of the basis, the Schr\"odinger equation
$$
 i \frac{\partial}{\partial t}\left(e^{i\alpha_{m,n}} \Vert n,m \rangle   \right)
= H e^{i\alpha_{n,m}}\Vert n,m\rangle
$$
reduces \cite{LewRie69} to the following equations for the
time-dependent coefficients $\alpha_{m,n}$
\begin{equation}\label{pha}
 \frac{d\alpha_{m,n}}{dt} = \langle n,m \Vert
 \left(i\frac{\partial}{\partial t} -H\right) \Vert n,m\rangle.
\end{equation}
One can thus split $\alpha_{m,n}$ into the sum
$\alpha_{m,n}=\alpha_{m,n}^g\!+\!\alpha_{m,n}^d$
of a geometric phase (including the Berry phase phenomenon
for slowly varying parameters)
\cite{Berry84,Sim83,AhaAna87} (see also \cite{MooShaWil89})
and a dynamical phase $\alpha^d_{m,n}$ respectively fulfilling
\begin{equation}\label{phagd}
 \dot\alpha^g_{m,n} = i\langle n,m \Vert
 \left(\frac{\partial}{\partial t} \Vert n,m\rangle\right),\qquad \qquad
\dot\alpha^d_{m,n} = -\langle n,m \Vert H \Vert n,m\rangle.
\end{equation}
To compute the matrix elements
$\langle n,m \Vert H \Vert n,m \rangle $, we express
$H$ in terms of $A_{\pm}, A_{\pm}^\dagger$. As a first step,

\begin{equation}\label{bb1}
\begin{array}{ll}
x_1 = \frac{-i}{\sqrt{2\kappa}}\sigma(a_1 -a_1^\dagger),
\qquad \quad & p_1 =\frac{-i}{\sqrt{2\kappa}}\dot\sigma\mu (a_1
-a_1^\dagger) - \frac{\sqrt{2\kappa}}{2\sigma}(a_1 +
a_1^\dagger),\\[8pt]
x_2 = \frac{-i}{\sqrt{2\kappa}}\sigma(a_2
-a_2^\dagger), \qquad \quad &
p_2 =\frac{-i}{\sqrt{2\kappa}}\dot\sigma\mu (a_2
-a_2^\dagger) - \frac{\sqrt{2\kappa}}{2\sigma}(a_2 + a_2^\dagger);
\end{array}
\end{equation}
in terms of $A_{-}, A_+^\dagger, A_{+},
A_-^\dagger$ we find
\begin{equation}
\begin{array}{l}
 x_1 = \frac{-i\sigma}{2\sqrt{\kappa}} \left( A_{-} \!-\!A_+^\dagger
\!+\! A_{+} \!-\! A_-^\dagger\right),\\[8pt]
p_1 = \frac{-i\dot\sigma\mu}{2\sqrt{\kappa}}\left(A_{-}
\!-\!A_+^\dagger \!+\!A_{+} \!-\! A_-^\dagger\right)
-\frac{\sqrt{\kappa}}{2\sigma}
\left(A_{-} \!+\! A_+^\dagger \!+\! A_{+} \!+\! A_-^\dagger\right),\\[8pt]
x_2 = \frac{\sigma}{2\sqrt{\kappa}}
\left(A_{-} \!-\! A_+^\dagger \!-\! A_{+} \!+\! A_-^\dagger\right),\\[8pt]
p_2 = \frac{\dot\sigma\mu}{2\sqrt{\kappa}}\left(A_{-} \!-\!A_+^\dagger
\!-\!A_{+} \!+\! A_-^\dagger\right) -i\frac{\sqrt{\kappa}}{2\sigma}
\left(A_{-} \!+\! A_+^\dagger \!-\! A_{+} \!-\!A_-^\dagger\right).
\end{array}
\end{equation}
and
\begin{equation}
\begin{array}{ll}
 x_1\!+\!ix_2 = \frac{i\sigma}{\sqrt{\kappa}}
\left(A_-^\dagger \!-\! A_{+} \right),\qquad &
p_1\!-\!ip_2 = \frac{i\dot\sigma\mu}{\sqrt{\kappa}}
\left(A_+^\dagger\!-\!A_{-}
\right)\!-\!\frac{\sqrt{\kappa}}{\sigma}
\left(A_{-} \!+\! A_+^\dagger \right),\\[8pt]
x_1\!-\!ix_2 = \frac{i\sigma}{\sqrt{\kappa}}
\left(A_+^\dagger \!-\! A_-\right),
\qquad  &
p_1\!+\!ip_2 = \frac{i\dot\sigma\mu}{\sqrt{\kappa}}
\left(A_-^\dagger
\!-\!A_{+} \right) \!-\!\frac{\sqrt{\kappa}}{\sigma}
\left(A_{+} \!+\!A_-^\dagger\right).
\end{array}
\end{equation}

Replacing these formulae in (\ref{hmt}) we find
\begin{eqnarray}
 H &=& \frac{1}{2\kappa}\!
\left(\dot\sigma^2\mu +\frac{\kappa^2}{\mu\sigma^2}
 +\omega^2\sigma^2 \!\right)\!
 \left(\!A_-^\dagger A_{-} + A_+^\dagger A_{+} +1 \!\right)\!-
 \nu\!\left(\!A_-^\dagger A_{-} -
A_+^\dagger A_{+}  \!\right)\\
&+&\left[\left(\frac{i\sigma\mu}{\sqrt{2\kappa}} +
\frac{\sqrt{\kappa}}{\sqrt{2}\sigma}\right)^2
\!\!-\frac{\omega^2\sigma^2}{2\kappa} \right]A_{-}A_{+} +
\left[\left(\frac{-i\sigma\mu}{\sqrt{2\kappa}} +
\frac{\sqrt{\kappa}}{\sqrt{2}\sigma}\right)^2
\!\!-\frac{\omega^2\sigma^2}{2\kappa}
\right]A_+^\dagger  A_-^\dagger.\nonumber
\end{eqnarray}
Only the operators in the first line
contribute to the matrix elements $\langle n,m \Vert H \Vert n,m \rangle $,
the ones in the second line lead to vanishing terms. So we finally find
\begin{eqnarray}
 \langle n,m \Vert H \Vert n,m \rangle = \frac{n+1}{2\kappa}
\left(\dot\sigma^2\mu +\frac{\kappa^2}{\mu\sigma^2}
 +\omega^2\sigma^2 \right)- m\nu.         \label{dyn}
\end{eqnarray}

Now we wish to evaluate
\begin{equation}\label{mat}
 \langle n,m \Vert \frac{\partial}{\partial t} \Vert n,m \rangle
= \mathcal{N} \langle n,m \Vert \frac{\partial}{\partial t} \left[
(A_-^\dagger)^{\frac{1}{2}(n+m) }
(A_+^\dagger)^{\frac{1}{2}(n-m)}\Vert 0,0 \rangle\right].
\end{equation}
$\Vert  0,0 \rangle$ is determined up to a time-independent phase.

On one hand, deriving (\ref{eigenvalues}) with $n\!=\!m\!=\!0$
with respect to $t$, we find
\begin{equation}
\left(\frac{\partial I}{\partial t}\right) \Vert  0,0 \rangle =
(2\kappa -I) \left(\frac{\partial}{\partial t}\Vert 0,0  \rangle\right),\qquad\qquad
L\left(\frac{\partial}{\partial t}\Vert 0,0  \rangle\right) =0. \label{der}
\end{equation}
As the left hand-side of (\ref{der})$_1$ is different from zero,
$\frac{\partial}{\partial t}\Vert 0,0  \rangle$ cannot vanish,
but must be annihilated by $L$. We determine it by realizing
 $ a_a,\Vert 0,0  \rangle$ in configuration space:
\begin{equation}
a_a= \frac{1}{\sqrt{2\kappa}}\left[
\left(\dot\sigma\mu \!+\!i\frac{\kappa}{\sigma} \right)x_a
+i\sigma \partial_a\right],\qquad\quad
\psi_0(x,t)=\frac {\kappa}{\sigma\sqrt{\pi}}
\exp\!\left[-\!\left(\frac {\kappa}{\sigma^2}\!-\!
\frac {i\dot\sigma\mu}{\sigma}\right)\!\frac {x\!\cdot\! x}2\right];
\label{config}
\end{equation}
in fact $\psi_0$ is annihilated by both $a_1,a_2$, therefore also by
$A_-,A_+$, and is normalized w.r.t. the scalar
product $(\psi,\psi')=\int\!d^2\!x \,\overline{\psi(x)} \psi'(x)$.
Deriving (\ref{config})$_2$ we find
$$
\frac{\partial\psi_0}{\partial t}(x)=
\left[\left(\frac{i\ddot\sigma\mu}{\sigma}
+ \frac{i\dot \mu\dot\sigma}{\sigma} -
\frac{i\dot\sigma^2\mu}{\sigma^2}+\frac{2\kappa\dot\sigma}{\sigma^3}
\right)\frac {x\!\cdot\! x}2-\frac{\dot\sigma}{\sigma}\right]\psi_0(x)
$$
By the Stokes theorem and some straightforward computations we obtain
$$
\int\!d^2\!x \,\partial_a(|\psi_0(x)|^2 x^a)=0
\qquad\Rightarrow\qquad
\int\!\!d^2\!x \,|\psi_0(x)|^2x\!\cdot\! x=\frac {\sigma^2}{\kappa}
\int\!\!d^2\!x \,|\psi_0(x)|^2=\frac {\sigma^2}{\kappa}.
$$
From the two previous equations we conclude
\begin{eqnarray}
\langle 0,0 \Vert\left(\frac{\partial}{\partial t}\Vert 0,0
\rangle\!\right) &=&\int\!\!d^2\!x \:\overline{\psi_0}
\,\frac{\partial\psi_0}{\partial t} =\int\!\!d^2\!x
\:\overline{\psi_0} \,\psi_0
\left[\left(\frac{i\ddot\sigma\mu}{\sigma} +\frac{i\dot
\mu\dot\sigma}{\sigma} -
\frac{i\dot\sigma^2\mu}{\sigma^2}+\frac{2\kappa\dot\sigma}{\sigma^3}
\right)\frac {x\!\cdot\! x}{2}-\frac{\dot\sigma}{\sigma}\right]
\nonumber\\
&=& \frac {i\mu}{2\kappa}\left(\sigma\ddot\sigma+
\frac{\dot \mu\sigma\dot\sigma}{\mu} -
\dot\sigma^2\right).   \label{conclude}
\end{eqnarray}
On the other hand,
\begin{eqnarray}
 \frac{\partial}{\partial t}
 \left[(A_-^\dagger)^{\frac{1}{2}(n+m) }
(A_+^\dagger)^{\frac{1}{2}(n-m)} \right]&=&
\frac{n+m}{2}\frac{\partial A_-^\dagger}{\partial t}
(A_-^\dagger)^{\frac{1}{2}(n+m-3) }
(A_+^\dagger)^{\frac{1}{2}(n-m)}\nonumber\\
&+& \frac{n-m}{2} (A_-^\dagger)^{\frac{1}{2}(n+m)} \frac{\partial
A_+^\dagger}{\partial t} (A_+^\dagger)^{\frac{1}{2}(n-m -2)};
\end{eqnarray}
but from  (\ref{fin0}) it follows
$$
 \frac{\partial A_-^\dagger}{\partial t}
=\frac{1}{\sqrt{2}}\frac{\partial a_{1}^\dagger}{\partial t} +
\frac{i}{\sqrt{2}}\frac{\partial a_{2}^\dagger}{\partial t},\qquad\qquad
\frac{\partial A_+^\dagger}{\partial t}
=\frac{1}{\sqrt{2}}\frac{\partial a_{1}^\dagger}{\partial t} -
\frac{i}{\sqrt{2}}\frac{\partial a_{2}^\dagger}{\partial t},
$$
and by equations (\ref{g1}),
\begin{eqnarray}
 \frac{\partial A_-^\dagger}{\partial t} &=&
\frac{1}{2\sqrt{\kappa}}\left[\left(\ddot\sigma\mu + \dot
\mu\dot\sigma +
i\frac{\kappa\dot\sigma}{\sigma^2}\right)(x_1+ix_2)
-\dot\sigma (p_1+ip_2)\right]\nonumber\\
&=&\frac{i\mu}{2\kappa}\left[\left(\sigma\ddot\sigma + \frac{\dot
\mu\dot\sigma}{\mu} -\dot\sigma^2\right)A_-^\dagger
-\left(\sigma\ddot\sigma + \frac{\dot
\mu\dot\sigma}{\mu} -\dot\sigma^2+\frac{2i\kappa\dot\sigma}
{\mu\sigma}\right)A_+\right],\label{deriv1}\\
 \frac{\partial A_+^\dagger}{\partial t} &=&
\frac{1}{2\sqrt{\kappa}}\left[\left(\ddot\sigma \mu +\dot
\mu\dot\sigma +
i\frac{\kappa\dot\sigma}{\sigma^2}\right)(x_1-ix_2)
-\dot\sigma (p_1-ip_2)\right]\nonumber\\
&=&\frac{i\mu}{2\kappa}\left[\left(\sigma\ddot\sigma + \frac{\dot
\mu\dot\sigma}{\mu} -\dot\sigma^2\right)A_+^\dagger
-\left(\sigma\ddot\sigma + \frac{\dot \mu\dot\sigma}{\mu} -
\dot\sigma^2+\frac{2i\kappa\dot\sigma}{\mu\sigma}\right) A_-\right].
\label{deriv2}
\end{eqnarray}
Inserting (\ref{conclude})-(\ref{deriv2}) into (\ref{mat})
and using (\ref{neq1}) we finally find
\begin{equation}
 \langle n,m \Vert \frac{\partial}{\partial t} \Vert n,m \rangle
=i \frac{n\!+\!1}{2\kappa}\left( \sigma\ddot\sigma\mu +
\dot \mu\sigma\dot\sigma - \dot\sigma^2\mu\right)=
i \frac{n\!+\!1}{2\kappa}\left( \frac{\kappa^2}{\mu\sigma^2} -
\omega^2\sigma^2-\dot\sigma^2\mu \right), \label{conclude2}
\end{equation}
as the operators $A_+,A_-$ give vanishing contributions to
the scalar product.
Replacing (\ref{dyn}) and  (\ref{conclude2}) in
 (\ref{pha}), the latter takes the form
\begin{equation}
 \frac{d\alpha_{n,m}}{dt} = -(n+1)\frac{\kappa}{\mu\sigma^2} +
m\nu
\end{equation}
(there is a partial cancellation of the geometric and dynamical phase
derivatives), which is solved by
\begin{equation}\label{phs}
\alpha_{n,m} (t)= -(n+1)\int^t \!\! dt'\:\frac{\kappa}{\mu(t')\sigma^2(t')} +
m\int^t \!\! dt'\:\nu(t').
\end{equation}

\section{Conclusions}
We have derived a class of exact invariants for the time-dependent, isotropic
two dimensional harmonic oscillator in a magnetic field, 
both on commutative and noncommutative plane. 
This applies also to the time-dependent
Landau problem (which is obtained choosing zero 
harmonic oscillator frequency
$\Omega(t)$). The angular momentum $L$ (absent in the one dimensional
case treated in Refs. \cite{lewis}) is trivially another invariant.
Using these two invariants, we
have constructed and an orthonormal basis of the associated Hilbert
space consisting of solutions of the  time-dependent Schr\"odinger
equation. The presence of the noncommutative parameter $\theta$
or of a time-dependent mass does
not change too much the results with respect to the commutative
space with constant mass, but induces a first order derivative term in the auxiliary
equation (\ref{neq1}), that is a general form of the Ermakov-Pinney
equation in Refs.\cite{PenYan07}. When we set $\theta=0$, then the equation
(\ref{phs}) are the same as the results in \cite{LewRie69}, but in two
dimensions. For some choices of $\Omega(t)$, exact solutions of
(\ref{neq1}) can be found and then the corresponding solutions of
the Schr\"odinger equation in close form. It would be also good to
calculate the transition amplitude connecting any initial state in
the remote past to any final state in the remote future with
constant $\omega$; we hope to report
on this point elsewhere.

{\it{Note added in proof}}. In the final stage of publication of the 
present paper Prof. Dodonov has brought to our attention the very 
interesting papers,\cite{Malkin}, \cite{dodonov}, where among 
other things exact invariants in the form of generalized creation 
and annihilation operators for a general N-dimensional harmonic 
oscillator were constructed.\\

\noindent{\bf Acknowledgments}

\noindent The work of one of us (L. G.) is supported by the
Associate and Federation Schemes and the HECAP Section of ICTP.


\begin{thebibliography}{10}
\bibitem{footnote}
Other t-dependent electric potentials have been considered in the 
literature, e.g., $\varphi(x) = e[E_1(t)x_1 + E_2(t)x_2]$ in Ref.\cite{Chee09}.
In Ref.\cite{Chee10}, the interaction of a spin with a $3$-dim magnetic field
$B(t)$ with a general t-dependence (i.e., possibly changing direction) has
been considered; this can be considered as first step in view of a Landau problem
in a general magnetic field of this kind.
\end{thebibliography}
\end{document}